\documentstyle[prl,tighten,aps,epsf,amstex,multicol,times,bbm]{revtex}
\begin{document}

\draft

\newcommand{\proofend}{\hfill\rule[0pt]{.2cm}{.2cm} }

\newcommand{\lb}[1]{\\\smallskip{\it (#1)}}

\title{Quantum and classical 
correlations
in quantum Brownian motion}

\author{Jens Eisert and Martin B.\ Plenio}
\address{QOLS, 
Blackett Laboratory, Imperial College of Science, Technology, and 
Medicine, London SW7 2BW, UK}

\date{\today}
\maketitle

\begin{abstract}
We investigate the entanglement properties of 
the joint state of a distinguished quantum system
and its environment in the quantum Brownian
motion model. This model is a frequent
starting point for investigations of environment-induced
superselection. Using recent methods from 
quantum information theory, we show that 
there exists a large class of initial states
for which no entanglement will be created at all 
times between the system of salient interest and the 
environment. If the distinguished system
has been initially prepared in a pure Gaussian 
state, then entanglement is created immediately,
regardless of the temperature of the environment and
the non-vanishing coupling.
\end{abstract}


\begin{multicols}{2}
\narrowtext
No quantum system is completely
isolated from its environment.
This basic yet fundamental
observation has been one of the key
insights allowing an appropriate understanding
of the dynamical emergence of classical properties in 
quantum systems. 
Not all initial states are 
equally fragile under the interaction
of a distinguished quantum 
system with its environment,
and a relatively
robust set of so-called preferred or
pointer states is selected dynamically,
a process typically referred to as 
environment-induced superselection
(einselection) or simply decoherence.
This process is thought to play 
an important role
in the transition from quantum to classical \cite{Dec,Dec2}.
The most frequently employed model in
investigations of einselection is the 
quantum Brownian motion model 
\cite{Caldeira,HUANDOTHERS}. 
In this model one considers a distinguished quantum
oscillator which is linearly coupled via the
position operators to an environment consisting
of many harmonic oscillators. 
Initially, the state of
the system of interest and its
environment are assumed to be 
uncorrelated, and the state of the environment is
taken to be the  canonical (Gibbs) state with
respect to some temperature.
The typical argument is that starting from the 
initial situation,
the product 
state of the composite quantum system
turns into a correlated state due to the interaction.
If one considers the reduced state of the 
distinguished system one finds 
that it undergoes 
dissipation and decoherence.
In the context of quantum Brownian motion it
is often argued that
entanglement is unavoidable.
%
%

It is the aim of this letter to
revisit the question of the
creation of entanglement in quantum Brownian
motion with recent powerful
methods from quantum 
information theory \cite{Intro1,Mention}. 
Our analysis will be split into two parts.
In the first part 
we will show that surprisingly,
quantum Brownian motion does not
necessarily create entanglement
between the distinguished system
and its environment. 
The joint state of the system and its environment
may be separable at all times, that is, not entangled 
\cite{Separable,Werner89}:
All correlations are merely 
classical in the sense that one could 
prepare the same state by mixing
product states, which can in turn be 
prepared by implementing local quantum 
operations only. By definition, separable states
do not violate any Bell inequality.
We then explicitly construct initial states with
the property that no entanglement is created:
they are mixed Gaussian states
which are nevertheless different from
Gibbs states. In contrast to
the finite-dimensional setting, where
a high degree of mixing automatically implies 
separability \cite{SB}, 
the existence of such initial states 
is not obvious \cite{MyInfi}.
The second part of our analysis is concerned
with the question whether there exist
initial states of the distinguished oscillator
for which the joint state becomes 
immediately entangled. This question will
be answered positively, and it
will be demonstrated that 
{\em all} pure Gaussian states have this property, regardless
of the initial temperature of the environment.

From now on 
the distinguished quantum oscillator
will be called $S$, 
the environment
will be referred to as $E$.
In the quantum Brownian motion model 
\cite{Caldeira,HUANDOTHERS},
the total Hamiltonian consists
of three parts
$H=H^S\otimes {\mathbbm{1}}+ {\mathbbm{1}}\otimes
H^E+H^I$, where
\begin{eqnarray}
	H^S&=&
\frac{1}{2m_1} P_1^2+
        \frac{m_1\omega_1^2}{2}X_1^2,\,\,\,\,
	H^I=- 
	X_1\otimes \sum_{j=2}^{N+1} \kappa_j X_j,
	\nonumber\\
	H^E&=&\sum_{j=2}^{N+1}
	\biggl(
        \frac{1}{2m_j} P_j^2+
        \frac{m_j\omega_j^2}{2}X_j^2
        \biggr).
\end{eqnarray}
The frequencies $\omega_1,\ldots,\omega_{N+1}$
and coupling constants $\kappa_2,\ldots ,\kappa_N$
are taken to be positive. 
For convenience, we set $\omega_{1}=1$,
all masses to be equal, $m_j=1$ for
$j=1, \ldots ,N+1$, and we
require that 
the $(N+1)\times (N+1)$-matrix $V$ 
corresponding
to the potential energy
is positive, 
where
$V_{1,1}=
\omega_1^2/2$ and
$V_{j,j}=\omega_j^2/2,$
$V_{1,j}=V_{j,1}=-\kappa_j$ for $j=2,...,N+1$,
and all other entries of $V$ are zero. 
Typically, one assumes product 
initial conditions \cite{Caldeira,HUANDOTHERS},
%
%
$\rho_{0}=\rho^S_{0}\otimes \rho^E_{0}$,
where the environment is initially in the 
Gibbs state
$\rho^E_{0}= \text{exp}(-\beta H^E)/\text{tr}[\text{exp}(-\beta H^E) ]$
associated with some inverse temperature $\beta$. 
This model together with the above additional
assumptions will be later referred to as QBM model 
in the more specific sense. 
The time evolution of the reduced state
with respect to $S$ can 
be determined without approximations
\cite{HUANDOTHERS}: for all
spectral densities $I(\omega)=\sum_{j=2}^{N+1} \kappa_{j}^{2}
\delta(\omega-\omega_{j})/(2\omega_{j})$
one can derive a differential equation
that specifies 
the dynamical map.
This completely positive
	map ${\cal E}_t$, $t\in[0,\infty)$, 
	maps an initial state $\rho_0^S$ of
	$S$ on the state 
	$\rho_t^S= {\cal E}_t (\rho_0^S)= 
	\text{tr}_{E}[U_{t}(\rho^{S}_{0}\otimes \rho^{E}_{0})U_t^{\dagger}]$ at 
	a later time $t$,  
	where $U_{t}:=\text{exp}(-i H t)$.

We will first
clarify the notation that will be used subsequently.
It will turn out to be
appropriate 
not to investigate the state 
on the infinite-dimensional 
Hilbert space
${\cal H}={\cal H}^{S}\otimes {\cal H}^{E}$
of the joint system
directly, but rather its associated covariance matrix.
Throughout the paper we will make  repeated
use of the formalism of covariance matrices
and their manipulation 
by means of symplectic transformations 
\cite{Wolf,Simon}. 
The $2n$ canonical self-adjoint 
operators corresponding to position and momentum
of a system with $n$ degrees of freedom
can be collected in a 
row vector ${\mathbf{O}}=(O_1,\ldots,O_{2n})=
(X_1,P_1,\ldots,X_{n},P_{n})$.
The canonical
commutation relations (CCR)
can then be written in matrix form as
$[O_j, O_k]=i(\Sigma_{2n})_{j,k}$, 
giving rise to the  
skew-symmetric block diagonal real 
$2n\times 2n$-matrix $\Sigma_{2n}$ (or simply
$\Sigma$ when the size of the matrix is clear
from the context).
Gaussian states, which are defined through their property that the
charateristic function is a Gaussian function in
phase space, can be characterized 
in a convenient way 
through their moments. 
The first moments
$\langle O_j\rangle_\rho$, $j=1,\ldots,2n$, 
are the expectation
values of the canonical coordinates. 
The $2n\times 2n$ covariance matrix $\Gamma$, 
\begin{eqnarray}
	\Gamma_{j,k}
	&=&
	2 \text{tr}\left[
	\rho \left(O_j-\langle O_j\rangle_{\rho} \right) 
	\left(O_k-\langle O_k\rangle_{\rho}  \right)
	\right]  - i \Sigma_{j,k},
\end{eqnarray}
satisfying the Heisenberg uncertainty principle
$\Gamma + i \Sigma\geq 0$,
embodies the second moments of a state $\rho$.
For later considerations
we give the set of covariance matrices of 
a system with $n$ degrees of freedom
the name
$C_{2n}:=\{ \Gamma\in M_{2n}: \Gamma= \Gamma^T, \Gamma + i \Sigma \geq 0\}$,
where
$M_{2n}$ is the set of real $2n\times 2n$-matrices. 
Particularly important will be 
covariance matrices of Gibbs 
states
$\exp(-\beta H)/\text{tr}[\exp(-\beta H)]$
with respect to a Hamiltonian $H$
and the 
inverse temperature $\beta$, which are important
examples of Gaussian states. 
For brevity, the corresponding covariance matrix will from 
now on be denoted 
as $\Gamma{(\beta H)}$.
We will frequently employ linear 
transformations
from one set of canonical coordinate to another
which preserve the CCR, meaning that $S\Sigma S^T=\Sigma$. 
Such transformations 
form the group of (real linear) 
symplectic transformations $Sp(2n ,\mathbbm{R})$.
A symplectic
transformation $S\in Sp(2n ,\mathbbm{R})$
results in a change of the covariance matrix
according to $\Gamma\longmapsto S \Gamma S^T$,
on the level of the states it is associated with
a unitary operation 
$\rho\longmapsto U(S)\rho U(S)^\dagger$.

We are now in the position to state the first Proposition. It 
is concerned with the fact that there exist initial states of 
the distinguished system and an initial temperature of the
environment such that the joint state stays separable for all
times. In the second part of the proof a lower bound for the
required 
inverse temperature of the bath will be determined.
\smallskip

\noindent {\bf Proposition 1.\ }
	{\it In the QBM model,
	for any choice of
	coupling constants 
	$(\kappa_2,\ldots,\kappa_{N+1}$),
	$\kappa_{j}\geq 0$,
	and any frequencies $(\omega_2,\ldots,\omega_{N+1})$,
	$\omega_{j}>0$,
	there exists a Gaussian initial state
	of the system $\rho^S_{0}$ with
	covariance matrix $\Gamma^{S}_{0}$
	and an
	inverse temperature
	$\beta>0$
	of the environment such that
	\begin{equation}
	\rho_{t}=U_{t}( \rho^S_{0} \otimes 
	\text{exp}(-\beta H^E) 
	/
	\text{tr}
	[\text{exp}(-\beta H^E)]) 
	U_{t}^{\dagger}
	\end{equation}
	is not entangled for all times $t \in[0,\infty)$.
	Let $\omega_{\infty}:=
	\max\{\omega_{2},\ldots,\omega_{N+1}\}
	$, $\delta:= 2\sum_{j=2}^{N+1}\kappa_{j}^{2}$, 
	and $\Omega:=(\omega_{\infty}^{2}+2 \delta^{1/2})^{1/2}$,
	then the above inverse temperature 
	is bounded from
	below by the smallest $\beta$ that satisfies
	$\Gamma_0^S 
	\oplus \Gamma{(\beta H^E)}\geq \Gamma{(\gamma H)}$,
	where $\gamma:=\min\{2,\log(1+2/\Omega)/\Omega\}$.
}

\smallskip

{\it Proof of Proposition 1.}\/
The system $S$
and environment $E$
together form a system with
$N+1$ canonical degrees of freedom, where now
${\mathbf{O}}=(O_1,\ldots,O_{2N+2})=
(X_1,P_1,X_2,P_2,\ldots,P_{N+1})$. There exists a $T \in Sp(2N+2,{\mathbbm{R}})$ such that
the Hamiltonian $\tilde H$ in the new canonical
coordinates
$\tilde{\mathbf{O}}^T
= T {\mathbf{O}}^T$
is the Hamiltonian
\begin{equation}
	\tilde H= \sum_{j=1}^{N+1} \frac{\tilde O_{2j-1}^2}{2}
	+ \sum_{j=1}^{N+1} \tilde\omega_j^2 
	\frac{\tilde O_{2j}^2}{2},
\end{equation}
of $N+1$ uncoupled oscillators,
with real numbers $\tilde \omega_j^{2}/2$, 
$j=1,\ldots,N+1$,
which are the eigenvalues of the positive
matrix $V$. Due to the fact that the coupling is
restricted to the coordinates associated with
positions, $T$ is both orthogonal and symplectic, i.e.,
$T\in Sp(2N+2,{\mathbbm{R}})\cap O(2N+2)$.
This symplectic transformation can be achieved 
by applying an appropriate orthogonal matrix $M\in SO(N+1)$
on the canonical variables corresponding to position,
$(\tilde O_1,\tilde O_3,\ldots,\tilde O_{2N+1})^T=
M (O_1,O_3,\ldots,O_{2N+1})^T$, such that $H$ becomes
diagonal in position, 
and the same matrix $M$ on the momentum variables, 
$(\tilde O_2,\tilde O_4,\ldots,\tilde O_{2N+2})^T=
M (O_2,O_4,\ldots,O_{2N+2})^T$. 
In these canonical coordinates 
the covariance matrix
corresponding to the Gibbs state
w.r.t.\ $\delta$
becomes 
\begin{eqnarray}\nonumber
	\Gamma{(\delta \tilde H)} &= &
	\text{diag}
	(\Gamma{(\delta \tilde H)}_{1,1},\ldots,
	\Gamma{(\delta \tilde H)}_{2N+2,2N+2}),\\	
	\Gamma{(\delta \tilde H)}_{2j-1,2j-1}&=&
	f(\delta \tilde \omega_j)/\tilde \omega_j
	,\,\,\,
	\Gamma{(\delta \tilde H)}_{2j,2j}=
	f(\delta \tilde \omega_j) \tilde \omega_j,\label{hereshow}
\end{eqnarray}
$j=1,\ldots,N+1$,
where the function $f:{\mathbbm{R}}^+\longrightarrow
{\mathbbm{R}}^+$ is defined as 
$f(x)= 1+ 2/(\text{exp}(x)-1)$.
Let $\gamma>0$ be defined as above. We now show that
\begin{equation}\label{gam}
	\Gamma{(\gamma \tilde H)}\geq
	{\mathbbm{1}}_{2N+2}.
\end{equation}
The largest eigenvalue
of $V$ is given by its operator norm \cite{Bhatia}
$\|V\|$.
By adding and subtracting the same term one obtains
$\|V\| \leq  
\text{max}\{
    {\omega_{1}^{2}/2},\omega_{2}^{2}/2,\ldots,
    \omega_{N+1}^{2}/2\}+
\| V- \text{diag}(\omega_{1}^{2}/2,\omega_{2}^{2}/2,\ldots,
\omega_{N+1}^{2}/2)\|$, giving rise to
$    \|V\| \leq \omega_{\infty}^{2}/2+
 \delta^{1/2}$.   
The value
of $\|V\|$ is related to the largest frequency $\tilde
\omega_{\infty}:=
\max\{ \tilde\omega_{j}, j=1,\ldots,N+1\}$ by
$\|V\|={\tilde\omega}_{\infty}^{2}/2$. On using Eq.\ (\ref{hereshow}), one
finds after a few steps 
that indeed
$\tilde\Gamma{(\gamma \tilde H)}\geq 
{\mathbbm{1}}_{2N+2}$.

In order to proceed, we need to invoke the concept
of partial transposition. It has
been shown in Ref.\ \cite{Wolf} that in a 
system consisting of one oscillator in a
system $S$ and $N$ oscillators in an environment 
$E$, a Gaussian state is separable if
and only if its partial transpose is a quantum
state.
By using the matrix 
${\Sigma}^{T_E}:= 
\Sigma_2\oplus (- \Sigma_{2N})$, the criterion
can also be written in the form that 
 a state with covariance matrix $\Gamma$ is
 separable if and only if
 $\Gamma + i \Sigma^{T_E}\geq 0$ \cite{Wolf}.
The next step is to see that with
$\gamma>0$ as defined 
in Eq.\ (\ref{gam})
\begin{equation}\label{tem}
T^T
\Gamma{(\gamma \tilde H)}
T
+ i  \Sigma^{T_E}\geq {\mathbbm{1}}_{2N+2} + i  \Sigma^{T_E}
\geq 0,
\end{equation}
since 
$\| i \Sigma^{T_E} \| = 1$ and $T\in SO(2N+2)$.
%
%
Equipped with these preparatory tools,
one can construct a class of product 
initial states such that
the joint state of the system and its environment is 
separable at all times. These product states are
of the form 
states $\rho_{0}^S\otimes \rho_{0}^E$, 
where  $\rho_{0}^E=
\text{exp}(-\beta H^E)/\text{tr}[\text{exp}(-\beta H^E)]$
with respect to a certain inverse temperature $\beta>0$.
$\rho_{0}^E$ is already a Gaussian state, and 
the state of the distinguished 
system $\rho_{0}^S$ is taken to be Gaussian as well.
On the level of covariance matrices such an initial state
is represented as $\Gamma_0=\Gamma_0^S \oplus
\Gamma{(\beta H^E)}$, that is,
\begin{eqnarray}\nonumber
\Gamma_0&=& \Gamma_0^S\oplus \text{diag}((\Gamma_0)_{3,3},\ldots, 
(\Gamma_0)_{2N+2,2N+2} ),\\
(\Gamma_{0})_{2j-1,2j-1}&=&
f(\beta \omega_j)/\omega_j,\,\,\,\,\,
(\Gamma_{0})_{2j,2j}=
f(\beta \omega_j) \omega_j,\label{form}
\end{eqnarray}
for $j=2,\ldots,N+1$, where $\Gamma_0^S\in C_{2}$.
The covariance matrix $\Gamma_0^S\in C_{2}$ 
and $\beta>0$
are now chosen to be such that
	$\Gamma_0 -  T^T \Gamma{(\gamma \tilde H)} T\geq 0$
holds. Such covariance matrices and 
inverse temperatures always exist. 
For a given covariance matrix
$T^T \Gamma{(\gamma \tilde H)} T$ 
one can always choose $\beta>0$
and $\Gamma_0^S\in C_{2}$ in such a way that
$\Gamma_0 - T^T  \Gamma{(\gamma \tilde H)} T$ 
is diagonally dominant 
\cite{Bhatia}.
Since it is a symmetric matrix, diagonal dominance 
implies that 
$\Gamma_0 - T^T   \Gamma{(\gamma \tilde H)} T$ 
is a positive matrix. From the definition of the covariance
matrix $\Gamma{(\gamma \tilde H)}$ one can infer
that $\Gamma_0 - T^T  \Gamma{(\gamma \tilde H)} T \geq 0$ 
is equivalent to the requirement 
	$\Gamma_0^S 
	\oplus \Gamma{(\beta H^E)}\geq \Gamma{(\gamma H)}$,
which is the inequality in Proposition 1 giving
rise to the lower bound for $\beta$.

Since the Hamiltonian $H$ is a
quadratic polynomial in the canonical coordinates, 
time evolution $\rho_{0}\longmapsto \rho_{t}= 
U_{t}\rho_{0} U_{t}^{\dagger}$ 
is effected 
by a symplectic transformation on the level of covariance
matrices. There exists a continuous map $t\in[0,\infty)
\longmapsto S_t \in Sp(2N+2,{\mathbbm{R}})$,
such that 
given a covariance matrix $\Gamma_0\in C_{2N+2}$ 
at time $t=0$,
the covariance matrix of $\rho_t$ becomes
$\Gamma_t
:=
S_t \Gamma_0	S_t^T$. The aim
is to show that 
this matrix corresponds to a 
separable state, i.e., $ \Gamma_t 
+ i   \Sigma^{T_E}\geq0$ for all times.
As  $\Gamma{(\gamma H)}$
is the covariance matrix of a Gibbs state 
with respect to $H$,
	$S_t
	\Gamma{(\gamma H)}
	S_t^T =
	\Gamma{(\gamma H)}$
for all $t\in[0,\infty)$, and hence,
$S_t \Gamma{(\gamma H)} S_t^T+ i 
	\Sigma^{T_E}\geq 0$. 
Moreover, the matrix
$S_t  \bigl(\Gamma_0 -  \Gamma{(\gamma H)}
\bigr) 
S_t^T$
is positive, because
$\Gamma_0 -  \Gamma{(\gamma H)}$
is positive. It follows that
\begin{eqnarray}\nonumber
    \Gamma_{t}  + i  
\Sigma^{T_E}=
  S_t \bigl(\Gamma_0 -  \Gamma{(\gamma H)}
    \bigr) 
	S_t^T 
	+
	\Gamma{(\gamma H)}
	+ i  \Sigma^{T_E}
	\geq 0\label{core}
\end{eqnarray}
for all $t\in[0,\infty)$. As the state was initially a Gaussian
state, it remains Gaussian under time evolution.
Having a positive partial transpose is equivalent
with being separable for systems where
one of the parts consists of only one oscillator, 
which means that we can conclude that $\Gamma_t$ 
corresponds to a separable state for all times.
\proofend

So we have shown that for these initial conditions, no
entanglement will be created at all times. At this
point,
a remark might be appropriate concerning the
inverse 
temperature $\beta$ in Proposition 1. 
The question 
of the behaviour of $\beta$ is particularly relevant
when one performs a continuum limit as is typically done
when deriving quantum master equations. In this context, 
it is of interest to see that a lower bound for $\beta$
can be found that is independent on the number of oscillators.
We consider a sequence of Hamiltonians $\{H_N\}_{N=1}^\infty$
of a joint system with an environment consisting of $N$ oscillators,
each equipped with a coupling constant $\kappa_j^{N}$ and a 
frequency $\omega_j^{N}$, $j=2,...,N+1$. Take for each $N$
an equidistant distribution of frequencies,
such that $\omega_j^{N}= (j-1) 
\omega_\infty /N$, where $\omega_\infty$ 
is the largest (cut-off)
frequency.
Concerning the spectral densities we only 
make the 
assumptions that $\kappa_j^{N} = \alpha_{N} 
(\omega_j^{N})^p$ with $p>0$ and
$\alpha_{N}>0$,
which covers the Ohmic ($p=1$), the subohmic ($p<1$) and
the supraohmic case ($p>1$). The refinement must in all 
instances be made such that 
$ \sum_{j=2}^{N+1} (\kappa_j^{N})^2
= 2 \int d\omega I(\omega)\omega$
remains constant. 
Then one can show that there exists a strictly positive
lower bound for $\beta$ as in Proposition 1 which is independent
of the number of oscillators $N$, and the limit
$N\rightarrow \infty$ may be performed. A detailed sketch of the
argument will be  given in footnote  \cite{Argument}.

In turn, having this observation in mind one may ask whether
there are initial states $\rho_{0}^S$ for which 
one can be sure that entanglement will 
be created immediately, 
no matter how weak the interaction is 
between the system and its environment, and 
given any possibly very high initial temperature
of the environment.
We shall see that there exist such states: 
all pure Gaussian states have this property. \smallskip

\noindent
{\bf Proposition 2.\ }
{\it
	In the QBM model, 
	for any initial
	pure Gaussian state $\rho^S_0$ of $S$, any
	coupling constants $(\kappa_2,\ldots,\kappa_{N+1})$,
	$\kappa_{j}\geq 0$,
	any frequencies $(\omega_1,\ldots,\omega_{N+1})$,
	$\omega_{j}>0$, and any
	$\beta>0$, 
	the state
	$\rho_{t}=U_t( \rho^S_{0} \otimes \text{exp}(-\beta H^E)/
	\text{tr}
	[\text{exp}(-\beta H^E)])U_{t}^{\dagger}$ 
	is entangled for all times $t \in(0,\varepsilon]$
for an appropriate $\varepsilon>0$.}

\smallskip

{\it Proof.}
A Gaussian 
state $\rho_{0}^S$ 
with a covariance matrix $\Gamma_{0}^S\in C_{2}$ 
is pure iff $(\Sigma_{2}
\Gamma)^{2}
=-{\mathbbm{1}}_{2}$.
The task is to find an $\varepsilon>0$ such that
${\cal E}_t( \rho^S_{0} \otimes \text{exp}(-\beta H^E)/
\text{tr}
[\text{exp}(-\beta H^E)])$ is entangled for all $t\in(0,\varepsilon]$.
Assume w.l.o.g.\
that  $\langle O_{j}\rangle_{\rho_{0}}=0$
for $j=1,2$, i.e., all first moments vanish
initially. Since $(d/dt)|_{t=0} \langle O_{j}\rangle_{\rho_{t}}=0$,
at $t=0$
the covariance matrix $\Gamma_{t}$ of
$\rho_{t}=U_{t}\rho_{0} U_t^\dagger$ 
satisfies 
\begin{equation}
    \frac{d}{dt} \Bigl|_{t=0}  
\Gamma_{t} = W^{T} \Gamma_{0}  W
\end{equation} 
with some $W\in M_{2N+2}$, which can be explicitly
evaluated by making use of 
$ i \text{tr}[O_{j} O_{k}( d \rho_{t}/d t)|_{t=0}]=
    \text{tr}[ O_{j} O_{k} [H,\rho_{0}] ]$, $j,k=1 ,\ldots, 2N+2$.
The state $\rho_{t}$ 
is entangled at a time 
$t>0$ if the reduced state with respect to the system 
$S$ and one of the oscillators of the environment 
$E$ is an
entangled state. The covariance matrix of $S$ and
the oscillators of $E$ with label $2$
is given by a real $4\times 4$ principal submatrix of
$\Gamma_{t}$: its entries are 
$(\Gamma_{t})_{j,k}$ with $j,k=1,\ldots,4$. 
This matrix will from now on be called $\gamma_{t}\in C_{4}$;
it can be written in block form as
\begin{equation}\label{fo}
	\gamma_t= 
	\left(
	\begin{array}{cc}
	A_t& C_t\\
	C_t^T& B_t
	\end{array}
	\right),
\end{equation}
where
$A_t,B_t,C_t\in M_{2}$. 
The statement that the reduced state of $S$ and
the oscillator with label $2$ is entangled is
equivalent to the statement that
$\gamma_t + i \Sigma^{T_E}$ is not positive, where now
$\Sigma^{T_E}=\Sigma_2\oplus (-\Sigma_2)$. This in turn
is equivalent with 
the smallest eigenvalue of $( i  \Sigma^{T_E} \gamma_t )^2$
being smaller than one 
(which means that one of
the symplectic eigenvalues of the partial transpose of 
$\gamma_t$ is smaller than one \cite{Negativity}).
Let the smallest eigenvalue of
$ (i \Sigma_4^{T_E} \gamma_t)^2$ 
be denoted as $\lambda_t$,
and let for brevity
$d_{t}:= \text{det}(A_t)+\text{det}(B_t)- 2  \text{det}(C_t)  $.
The smallest eigenvalue $\lambda_t$
can then be expressed as 
	$\lambda_t=
	d_{t}/2
	- (
	d_{t}^2 /4
	- \text{det}(\gamma_t)
	)^{1/2}$ \cite{Negativity}.
At $t=0$, the covariance matrix $\gamma_0$
is of the form of Eq.\ (\ref{fo}) with
$A_0\in C_{2}$, $C_0=0$, and $B_0=\text{diag}(
f(\beta\omega_{2})/\omega_{2},f(\beta\omega_{2})\omega_{2})$, 
the latter matrix
satisfying $\text{det}(B_0)>1$ by definition.
As $S$ is initially in a pure state, $\text{det}(A_0)=1$.
Therefore, 
$\lambda_0=1$.
The first derivatives of $A_t$, $B_t$, and $C_t$ at $t=0$
can 
be computed 
to be given by
\begin{eqnarray}\nonumber
\frac{d}{dt}\Bigl|_{t=0} A_t &=& 
	\left(
	\begin{array}{cc}
	0& (A_0)_{2,2} -(A_0)_{1,1} \\
	(A_0)_{2,2} -(A_0)_{1,1} & - 2 (A_0)_{1,2}
	\end{array}
	\right),\\
\frac{d}{dt}\Bigl|_{t=0} C_t &=& \nonumber
	\left(
	\begin{array}{cc}
	0& - \kappa_2 (A_0)_{1,1} \\
	0&\kappa_2 (A_0)_{1,2}
	\end{array}
	\right),
	\\
\frac{d}{dt}\Bigl|_{t=0} B_t &=& 
	\left(
	\begin{array}{cc}
	0&  (B_0)_{2,2}-(B_0)_{1,1} \\
	(B_0)_{2,2}- (B_0)_{1,1} & 0
	\end{array}
	\right).
\end{eqnarray}
Hence, it follows that
	$(d \lambda_t /dt)|_{t=0}<0$.
This means that there exists an $\varepsilon>0$ such that
$\lambda_t<1$
for all $t\in(0,\varepsilon]$, which implies that
the partial transpose of the state corresponding to
$\gamma_t$ is not a state. 
We can conclude that the state $\rho_{t}$
associated with the covariance matrix $\Gamma_{t}$ is an entangled
state \cite{Vlatko}.\proofend

To summarize, we have investigated the entanglement
properties of the joint state of a 
distinguished system and
its environment in quantum Brownian motion. 
Surprisingly indeed, 
we found that there exists a large set of 
initial states of the system for which no entanglement is
created at all times. Also, we have shown that for
pure initial Gaussian states of the distinguished system
entanglement will be immediately created.
The tools we used
were mostly taken from the field of 
quantum information theory. In fact, we
hope that this letter can contribute to the line
of thought of applying methods from quantum 
information theory to the issue of 
emerging classicality in quantum physics.

We would like to thank P.\ H{\"a}nggi,
P.\ Talkner, C.\ Simon, H.\ Wiseman,
K.J.\ Bostr{\"o}m,
J.\ Kempe,   V.\ Scarani, J.R.\ Anglin, and
W.H.\ Zurek 
for inspiring discussions.
This work has been supported by the EU
(EQUIP, IST-1999-11053), 
the A.v.-Humboldt-Foundation,
and the EPSRC.

\end{multicols}


\begin{references}
\bibitem{Separable}
        A state $\rho$ of a bi-partite system with Hilbert space 
        ${\cal H}={\cal H}^{S}\otimes {\cal H}^{E}$
        is called separable \cite{Werner89}, 
        if it can be written
        as $\rho=\sum_{i} p_{i} 
        |\psi^{S}_{i}\rangle\langle \psi^{S}_{i}|\otimes
        |\psi^{E}_{i}\rangle\langle \psi^{E}_{i}|$
        where $\psi^{S}_{i}\in{\cal H}^{S}$,
        $\psi^{E}_{i}\in{\cal H}^{E}$ for all $i$,
        and $\{p_{i}\}_{i}$ forms a probability
        distribution. Classically correlated will be used
	synonymously with separable (however, see also Ref.\ \cite{ZurekCorr}).
\bibitem{Argument} 
The goal is to show that there exists a 
$\beta>0$ and a $\Gamma_0^S\in C_2$
such that $\Gamma_0^S\oplus \Gamma(\beta H^E_N)\geq
\Gamma(\gamma H_N)$ for all $N\in {\mathbbm{N}}$.
This means that 
the continuum limit may be 
performed, and the temperature of the environment that has
to be chosen in order to satisfy the assumptions of Proposition 1
does not diverge when performing the limit $N\rightarrow \infty$:
(i) The first step is to observe that it is sufficient to
prove that $\Gamma_0^S\oplus \Gamma(\beta H^E_N)\geq 2 (A_N\oplus
B_N)$,
where $A_N$ 
is the $2\times 2$ upper leading principle
submatrix of $\Gamma(\gamma H_N)$, and $B_N$
is the $2N\times 2N$ lower principle
submatrix of $\Gamma(\gamma H_N)$. 
This is sufficient, as if the latter is true, 
then
$\Gamma_0^S\oplus \Gamma(\beta H^E_N)- \Gamma(\gamma H_N)
= \Gamma_0^S\oplus \Gamma(\beta H^E_N)- 2(A_N\oplus B_N) + 
P \Gamma(\gamma H_N) P\geq 0$,
 where $P:=\text{diag}(-1,-1, 1,...,1)$.
(ii) The next step is to show that there exists a number $q>0$ such that
$2 \Gamma(\gamma H_N^E)+q {\mathbbm{1}}\geq B_N$
for all $N\in {\mathbbm{N}}$. (iv) There always exists
a $\beta>0$ such that $f(\omega \beta)\omega \leq 2 f(\omega\gamma)\omega +q$
and $f(\omega \beta)/\omega \leq 2f(\omega\gamma)/\omega +q$
for all $N\in {\mathbbm{N}}$ and all $\omega\in [0,\omega_\infty]$, 
and therefore, also
 $\Gamma(\beta H_N^E) \geq 2 \Gamma(\gamma H_N^E)+q  
{\mathbbm{1}}$ holds
for all $N\in {\mathbbm{N}}$. (v) The last step is
to see that there exists a $\Gamma_0^S\in C_2$ such that
$\Gamma_0^S\geq A_N$ for all $N\in {\mathbbm{N}}$. 
This is however obvious from the requirement that
$\sum_{j=2}^{N+1}(\kappa_j^N)^2$ is kept constant. 
Physically, it means that
the covariance matrix of the
reduced state of system $S$
in the Gibbs state
at inverse temperature $\gamma$ is not divergent in the continuum
limit.
\end{references}
\end{document}